\documentclass[aps,prl,twocolumn,superscriptaddress,nofootinbib,floatfix]{revtex4-2}

\usepackage{amsmath,amssymb}
\usepackage{mathrsfs}
\usepackage{graphicx}
\usepackage[colorlinks=true,linkcolor=blue,citecolor=blue,urlcolor=blue]{hyperref}
\usepackage{orcidlink}

\newcommand{\MP}{M_{\rm P}}
\newcommand{\Lq}{L^{1/4}}
\newcommand{\Treh}{T_{\rm reh}}
\newcommand{\chat}{\hat c_0}
\newcommand{\Dphi}{\Delta\phi}
\newcommand{\eperp}{\eta_\perp}
\newcommand{\shat}{\hat s}
\newcommand{\xip}{\xi_\phi}

\begin{document}

\title{Superheavy gravitational dark matter from runaway bubble collisions}

\author{Cheng-Wei Chiang\,\orcidlink{0000-0003-1716-0169}}
\email{chengwei@phys.ntu.edu.tw}
\affiliation{Department of Physics and Center for Theoretical Physics, National Taiwan University, Taipei 10617, Taiwan}
\affiliation{Physics Division, National Center for Theoretical Sciences, National Taiwan University, Taipei 106319, Taiwan}
\author{Van Que Tran\,\orcidlink{0000-0003-4643-4050}}
\email{vqtran@phys.ncts.ntu.edu.tw}
\affiliation{Physics Division, National Center for Theoretical Sciences, National Taiwan University, Taipei 106319, Taiwan}
\affiliation{Phenikaa Institute for Advanced Study, Phenikaa University, Nguyen Trac, Duong Noi, Hanoi 100000, Vietnam}

\begin{abstract}
A stable particle that couples only through gravity is pair-produced in collisions of runaway bubble walls by graviton exchange. For a transition of duration $1/\beta$, a Dirac fermion reaches the yield $Y_X\simeq6.8\times10^{-3}(H/\beta)(T_{\rm reh}/M_{\rm P})^3$, where $H$ is the Hubble rate of the vacuum energy, $T_{\rm reh}$ the reheating temperature and $M_{\rm P}$ the reduced Planck mass. The yield is flat up to masses near $(H/\beta)M_{\rm P}/\nu^2$, where $\nu$ of order one sets the wall-field excursion. Gravity alone then gives dark matter up to about $10^{17}$~GeV at reheating temperatures far below its mass, where freeze-in fails.
\end{abstract}

\maketitle

\emph{Introduction.}---So far, all evidence for dark matter (DM) is gravitational. The most economical possibility is a stable particle $X$ of mass $M_X$ that interacts with ordinary matter only through gravity, so that its abundance must also come from gravity. The known channels are limited by the energy available. Thermal freeze-in through graviton exchange in the Standard Model plasma~\cite{Garny:2015sjg,Garny:2017kha,Tang:2017hvq,Bernal:2018qlk} gives a yield $\propto(T_{\max}/\MP)^3$ that is Boltzmann suppressed for $M_X\gtrsim T_{\max}$. Here $T_{\max}$ is the highest plasma temperature and $\MP=(8\pi G)^{-1/2}\simeq2.4\times10^{18}$~GeV is the reduced Planck mass. Production through the inflaton condensate~\cite{Mambrini:2021zpp,Clery:2021bwz,Clery:2022wib} and at the end of inflation~\cite{Chung:1998zb,Ema:2018ucl,Kolb:2023ydq} reaches the inflaton mass or the inflationary Hubble rate $H_{\rm inf}$, which the absence of primordial tensor modes in the cosmic microwave background (CMB) bounds~\cite{BICEP:2021xfz}.

First-order phase transitions in which the bubble walls run away provide much larger energies. The latent heat goes into the walls, whose Lorentz factor $\gamma$ grows with the bubble radius. Colliding walls meet at center-of-mass energies up to about $2\gamma m$, where $1/m$ is the wall thickness. Through direct couplings these collisions produce particles far heavier than the temperature~\cite{Watkins:1991zt,Chung:1998ua,Kolb:1998ki,Falkowski:2012fb,Azatov:2021ifm,Baldes:2023cih,Giudice:2024tcp,Shakya:2024kmb,Shakya:2025qpi,An:2026sdu,Cheng:2026npt}, and graviton emission accompanies this production~\cite{Chen:2026beh}. Reference~\cite{Ghoshal:2026pew} describes the walls as beams of quasi-real quanta of the wall field and writes the DM abundance through the number of production events per unit wall area. References~\cite{Redi:2026nzi,Garani:2025qnm} express gravitational production by any sub-horizon source through its stress-tensor correlator. Reference~\cite{Redi:2026nzi} applies this to phase transitions and estimates the production by the part of the source that varies on the duration $1/\beta$ of the transition. For light particles, this estimate [Eq.~(3.18) in Ref.~\cite{Redi:2026nzi}] is proportional to $(H/\beta)(\Treh/\MP)^3$ when the gravitational-wave density of the envelope approximation is used. Here we compute the production by the hard part of the source, the colliding walls. For Dirac-fermion DM, it is about thirty times larger than that estimate, and it extends to masses far above $\beta$
(see Sec.~S4 of the Supplemental Material). 
The field quanta of two colliding walls annihilate through an $s$-channel graviton into $X$ pairs, a production channel that, to our knowledge, has not been studied.

\begin{figure}[t]
\centering
\includegraphics[width=\columnwidth]{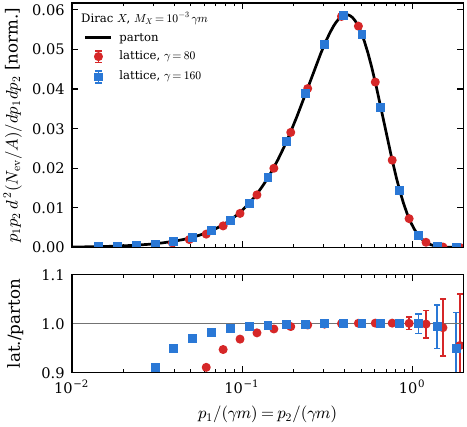}
\caption{Pair production in the collision of a $\phi^4$ kink and antikink in $1+1$ dimensions, along $p_1=p_2$, for a Dirac $X$ with $M_X=10^{-3}\gamma m$. Upper panel: the integrand of Eq.~\eqref{eq:NA} times $p_1p_2$, in units of $4c_X(0)\Dphi^4\gamma^2m^2/(\pi^2\MP^4)$. Curve: parton prediction. Points: lattice simulation at $\gamma=80$ (circles) and $160$ (squares), extrapolated to zero lattice spacing. Lower panel: ratio of lattice to parton. Error bars: numerical uncertainty of the lattice.  
}
\label{fig:lattice}
\end{figure}

We find that runaway bubble collisions produce every stable particle that couples to the transition sector only through gravity, with a yield given in closed form in the massless limit. For a given transition, this yield, corrected for the expansion, is a minimum abundance. Apart from percent-level gravitational contact terms and strong-gravity effects, only corrections near the highest masses, or entropy produced afterward, can lower it. 
Three results lead there. First, below the collision energy, the number of pairs produced per unit wall area depends only on the energies of the two walls for any wall profile. Second, for exponential nucleation, the sum over first collisions, oblique ones included, is exact in flat space. The yield is flat in $M_X$ up to masses set by $(H/\beta)\MP$, where $H/\beta$ is the duration of the transition in Hubble times. This holds for any reheating temperature $\Treh$, and the yield is about $1.5$ times lower than the estimate that assumes a single bubble size and head-on collisions. Third, nonlinear simulations of head-on wall collisions in $1+1$ dimensions confirm that the walls pass through each other where heavy pairs form. At $\gamma=80$ and $160$, the simulated pair production agrees with our analytic prediction at the percent level (Fig.~\ref{fig:lattice}).

\emph{Cross sections.}---The wall field $\phi$ is a real scalar whose potential has a false and a true vacuum separated by $\Dphi$, and $m$ is its mass in the true vacuum. Both $\phi$ and $X$ couple to the canonically normalized graviton through their stress tensors, $\mathcal L=-h_{\mu\nu}(T^{\mu\nu}_\phi+T^{\mu\nu}_X)/\MP$. The coupling has mass dimension $-1$, so the cross section of $\phi\phi\to h^*\to X\bar X$ grows with the squared center-of-mass energy $\shat$~\cite{Garny:2017kha,Tang:2017hvq,Mambrini:2021zpp},
\begin{equation}
\sigma=c_X(\varrho)\,\frac{\shat}{\MP^4}, 
\label{eq:xsec}
\end{equation}
where $c_X(0)={\chat}/{(960\pi)}$ and $\varrho=M_X^2/\shat$. A Dirac fermion has $\chat=1$. For a real scalar and a Proca field, $\chat$ depends on the curvature couplings.
The Dirac value does not, because curvature couplings change only the trace of the stress tensors, and a massless fermion does not couple to the trace. The massless Dirac yield therefore contains no gravitational parameter other than $\MP$, and we take it as our benchmark. Near the collision scale, the curvature coupling $\xi_\phi$ of the wall field enters, and we show $\xi_\phi=0$. 

\emph{One collision.}---A planar wall with field jump $\Dphi$, moving with $\gamma\gg1$, is a coherent state of massless $\phi$ quanta, which we call partons~\cite{Ghoshal:2026pew}. Their number per unit wall area $A$ and momentum $p$ is $dN/(A\,dp)=|F(p/\gamma)|^2/(\pi p)$, where $F$ is the Fourier transform of the gradient of the rest-frame profile. Partons of momenta $p_1$ and $p_2$ from the two walls collide head-on with $\shat=4p_1p_2$, and the number of production events per unit area is
\begin{equation}
\frac{N_{\rm ev}}{A}=\int_0^\infty dp_1\,dp_2\,\frac{dN_1}{A\,dp_1}\frac{dN_2}{A\,dp_2}\,\sigma(4p_1p_2).
\label{eq:NA}
\end{equation}
Since $\sigma$ is proportional to $p_1p_2$ for $M_X\ll\gamma m$, the double integral splits into two integrals of the parton density weighted by momentum. Each of them is the energy per unit area $E_i/A$ of one wall. For any mass, we write
\begin{equation}
\frac{N_{\rm ev}}{A}=\frac{4c_X(0)}{\MP^4}\,\frac{E_1}{A}\,\frac{E_2}{A}\,g(\mu),
\label{eq:theorem}
\end{equation}
where $\mu=M_X/(\gamma m)$ and $g(0)=1$ for every wall profile.  The profile enters only through the mass suppression $g$. At fixed wall energy and field excursion, $g$ falls to $0.1$ at nearly the same mass for $\tanh$, sine-Gordon and error-function walls.
We have derived the same result exactly at order $\MP^{-4}$ by treating the colliding walls as a classical source in the weak-field formula of Refs.~\cite{Zeldovich:1977vgo,Frieman:1985fr}. A single uniformly moving wall produces nothing~\cite{Gibbons:1975jb}, and the cross term of the stress tensor, $\propto\phi_1'(t+x)\,\phi_2'(t-x)$, reproduces Eq.~\eqref{eq:NA} with no additional coherence factor.

Both derivations assume that the walls pass through each other with superposed profiles~\cite{Strumia:2026uvz}.  Our simulations of a kink--antikink collision confirm free passage where heavy pairs form. Figure~\ref{fig:lattice} compares the integrand of Eq.~\eqref{eq:NA} with the lattice at $\gamma=80$ and $160$.
 For $p_1=p_2>0.2\gamma m$, the region that gives about $80\%$ of the yield along the diagonal, the lattice reproduces it within $0.3\%$ up to $0.7\gamma m$, and beyond that within the numerical uncertainty, which reaches $1$ to $2\%$ at $\gamma m$. At momenta of a few $m$, the potential energy of the overlapping walls, dropped by the parton picture, lowers the lattice result by about $2(m/p)^2$, a correction that moves to smaller $p/(\gamma m)$ as $\gamma$ grows. Between $0.1\gamma m$ and $\gamma m$ the lattice departs from the superposed walls by at most $0.5\%$ at $\gamma=80$ and $0.1\%$ at $\gamma=160$, consistent with the expected $1/\gamma^2$ corrections~\cite{Strumia:2026uvz}. Integrated over the simulated momenta, the lattice yield at both $\gamma$ agrees with the parton yield within a few percent up to $M_X=\gamma m$. These corrections fall with $\gamma$, so we expect them to be far smaller in the collisions of interest, where $\gamma\simeq\bar M_{\rm coll}/m\gtrsim10^3$ for $m\approx\Lq$. 
 Walls accelerated from rest and oblique collisions remain to be simulated.

\emph{Bubble population.}---In three dimensions, the two walls that meet at a point are not antiparallel. Let $\eperp=\sin(\theta/2)$, with $\theta$ the angle between their directions of motion. Partons then collide with $\shat=4p_1p_2\eperp^2$ and M{\o}ller velocity $2\eperp^2$, and the time integral of the two sheets gives $1/(2\eperp)$ per unit area of the collision surface. A patch of that surface therefore yields Eq.~\eqref{eq:theorem} times $\eperp^3$, with the collision energy scale $M_{\rm coll}=\sqrt{\gamma_1\gamma_2}\,m\,\eperp$ in place of $\gamma m$, where $\gamma_{1,2}$ are the Lorentz factors of the two walls.

We take a runaway transition in vacuum with nucleation rate $\Gamma\propto e^{\beta t}$ and $\beta\gg H$. The walls carry the latent heat $L$, so a wall of radius $r$ has $E/A=Lr/3$, and the released energy thermalizes into $g_*=106.75$ degrees of freedom at $\Treh$, with $L=(\pi^2g_*/30)\Treh^4$. Counting first collisions, where two walls reach a point before any other, the yield is proportional to
\begin{equation}
Q_\beta=\frac{1}{\beta V}\int dA_{\rm locus}\,\hat e_1\hat e_2\,\eperp^3,\qquad\hat e_i=\frac{\beta r_i}{3},
\label{eq:Qbeta}
\end{equation}
an integral over the interfaces of a Johnson--Mehl tessellation in a volume $V$, with $A_{\rm locus}$ their area. For exponential nucleation, the two radii and the angle are statistically independent on these interfaces~\cite{Moller:1992,Miles:1972}. Each $\beta r_i$ follows a Gamma distribution of shape three, so $\langle\hat e_i\rangle=1$. The angle gives $\langle\eperp^3\rangle=1/2$, and the interface area per volume is $2\beta/3$. Hence $Q_\beta=1/3$ exactly for first collisions in flat space, and a Monte Carlo of the first-collision surfaces gives $0.336\pm0.002$.
Oblique collisions halve the yield, and the spread of bubble radii raises it by a factor of $1.4$. Together they make the yield smaller by a factor of about $1.5$ than the estimate of Ref.~\cite{Ghoshal:2026pew}, which takes spheres of a single radius $(8\pi)^{1/3}/\beta$ and head-on collisions. With two particles per event, our main result is
\begin{equation}
Y_X=\frac{n_X}{s}=\frac{\chat}{480}\sqrt{\frac{g_*}{10}}\,\frac{H}{\beta}\left(\frac{\Treh}{\MP}\right)^3R(M_X),
\label{eq:Yuniv}
\end{equation}
where $n_X$ is the number density of $X$, $s$ the entropy density, $H^2=L/(3\MP^2)$ the Hubble rate of the vacuum energy, and $\sqrt{g_*/10}/480\simeq6.8\times10^{-3}$. Here $R\le1$ is the average of $g$ over the collisions, weighted by the yield. For first collisions and free passage in flat space, the coefficient is exact. On the expanding background, a Monte Carlo of the first-collision surfaces gives a yield lower than Eq.~\eqref{eq:Yuniv} by a factor close to $e^{-4H/\beta}$, from $0.67$ at $H/\beta=0.1$ to $0.95$ at $0.01$, at fixed latent heat. Section~S4 of the Supplemental Material gives this correction and the function $R(M_X)$.

A direct coupling of $X$ to the wall field gives a yield that falls with $M_X$~\cite{An:2026sdu,Ghoshal:2026pew}, so the flat yield is special to gravity.

The mass reach follows from energy conservation, $\sigma_{\rm wall}\gamma_i=Lr_i/3$, with the wall tension $\sigma_{\rm wall}=m\Dphi^2/6$. Averaged over the collisions with the yield as weight, the collision energy scale is
\begin{equation}
\bar M_{\rm coll}=\frac{525\pi}{256}\,\frac{L}{\beta\Dphi^2}\simeq11\,\frac{H}{\beta}\,\frac{\MP}{\nu^2}.
\label{eq:Mbar}
\end{equation}
The last form uses $H^2=L/(3\MP^2)$ and $\Dphi=\nu\Lq$, where $\nu$ is of order one for a potential with a single scale. The yield is halved at $M_X\simeq\bar M_{\rm coll}/6$ and falls to a tenth at $M_X\simeq0.45\,\bar M_{\rm coll}$. The reach is set by the duration of the transition and by the Planck mass, \emph{not by $\Treh$}. 
Because $\sigma\propto\shat$, 
the yield can be trusted only below the scale where the gravitational effective theory fails. We discard all yield from parton collisions above the species scale of the Standard Model, $\Lambda_{\rm cut}=0.46\,\MP$~\cite{Dvali:2007hz}. Its numerical factor is not fixed, so the highest masses we quote depend on the ultraviolet completion of gravity.

\begin{figure}[t]
\centering
\includegraphics[width=\columnwidth]{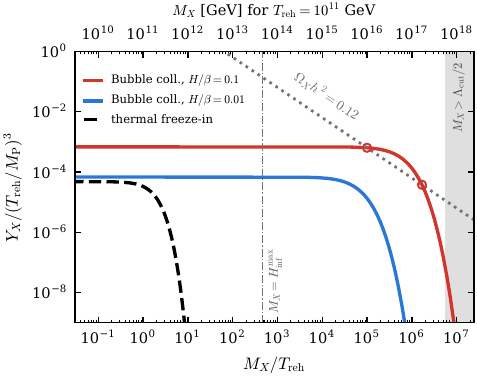}
\caption{Yield of Dirac-fermion DM that couples only through gravity, in units of $(\Treh/\MP)^3$, against $M_X/\Treh$ for the reheating temperature $\Treh=10^{11}$~GeV. 
Solid: 
first collisions of runaway bubbles,
Eq.~(\ref{eq:Yuniv}), with $\Dphi=3\Lq$ and $\xi_\phi=0$, for two values of $H/\beta$. Dashed: thermal freeze-in with instantaneous reheating. Dotted: the yield that gives the observed DM density, with circles at its crossings with the curve for $H/\beta=0.1$.
Dash-dotted: $M_X=H_{\rm inf}^{\max}$, the CMB bound on the inflationary Hubble rate. Gray: $M_X>\Lambda_{\rm cut}/2$, with $\Lambda_{\rm cut}$ the gravitational cutoff.}
\label{fig:money}
\end{figure}

\begin{figure}[t]
\centering
\includegraphics[width=\columnwidth]{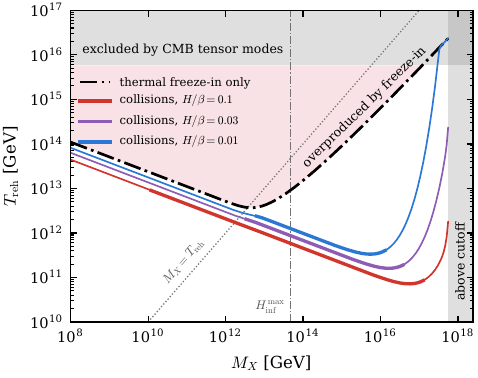}
\caption{Reheating temperature that gives the observed DM density for Dirac-fermion DM. Dash-dotted: thermal freeze-in only, with instantaneous reheating. Solid: first collisions as in Fig.~\ref{fig:money}, plus freeze-in, for three values of $H/\beta$. Thick segments mark the core of each curve, where the conditions of Sec.~S5 of the Supplemental Material hold. Pink: overproduced by freeze-in. Gray: reheating temperatures excluded by the CMB bound on tensor modes, or $M_X>\Lambda_{\rm cut}/2$. Dotted: $M_X=\Treh$. Vertical dash-dotted: $M_X=H_{\rm inf}^{\max}$.}
\label{fig:plane}
\end{figure}

\emph{Dark matter.}---Graviton exchange in the plasma after reheating also produces $X$. 
For instantaneous reheating and $M_X\ll\Treh$ it gives $Y_{\rm FI}=4.8\times10^{-5}(\Treh/\MP)^3$ for a Dirac fermion, 
which we evaluate following Ref.~\cite{Redi:2026nzi}.  Bubble collisions exceed it by the factor $142\,H/\beta$ for light $X$ and by an exponentially growing factor above $\Treh$. 
Figure~\ref{fig:money} compares the channels at one reheating temperature. The thermal yield collapses once $M_X$ exceeds a few times $\Treh$, while the collision yield stays at Eq.~\eqref{eq:Yuniv} until $M_X$ approaches $\bar M_{\rm coll}$. 
Production at the end of inflation 
falls steeply above $H_{\rm inf}$,
which the CMB bounds by $H_{\rm inf}^{\max}\simeq5\times10^{13}$~GeV~\cite{BICEP:2021xfz}. For $H/\beta=0.1$ the collision yield stays flat up to about $10^5\,\Treh$, where it crosses the dotted line and gives the DM mass at this $\Treh$. The second crossing depends on the tail of the wall profile.

Figure~\ref{fig:plane} shows the reheating temperature at which each channel gives the observed abundance, $\Omega_Xh^2=0.12$~\cite{Planck:2018vyg}. Freeze-in alone, with instantaneous reheating, needs $\Treh$ above about $4\times10^{12}$~GeV at every mass, and the region above its line is overproduced. Slower reheating heats the plasma above $\Treh$ and extends freeze-in to heavier masses, but the CMB bound still caps its reach near $10^{17}$~GeV. Bubble collisions open a large part of the region below that line. 
 Along each bubble-collision curve, $\Treh$ falls as $M_X^{-1/3}$ until $M_X$ approaches $\bar M_{\rm coll}$ and the curve turns up. On the cores of these curves, first collisions in a runaway transition give all of the DM for $M_X$ between $10^{10}$ and $3\times10^{17}$~GeV, for $H/\beta$ between $0.01$ and $0.1$ 
 and $\nu$ between two and three (Fig.~\ref{fig:plane} shows $\nu=3$). Gravity alone therefore produces DM at reheating temperatures up to six orders of magnitude below its mass, beyond the reach of freeze-in and, for masses above the inflaton mass and $H_{\rm inf}$, beyond that of inflationary production. The heaviest of these masses rely on collision energies near $\Lambda_{\rm cut}$. The expansion correction, not included, raises the curves by at most $14\%$ in $\Treh$. For the estimate of Ref.~\cite{Ema:2019yrd}, production at the end of inflation adds less than $10\%$ above $M_X\approx2\times10^{15}$~GeV for any $H_{\rm inf}$ allowed by the CMB, and below that mass it requires $H_{\rm inf}$ below a few percent of $M_X$. For an abrupt end of inflation, our Bogoliubov calculation moves this mass into the range $3\times10^{15}$ to $8\times10^{15}$~GeV. 

\emph{A realization.}---Runaway walls arise in the scenario of Ref.~\cite{An:2026ghw}, where the hot Big Bang begins with the decay of a false Standard Model vacuum after inflation. Unlike our benchmark, its nucleation rate is constant and its vacuum energy is only part of the total density. A cosmological Monte Carlo on the self-consistent background shows that Eq.~\eqref{eq:Yuniv} still gives the yield at fixed latent heat to about $10\%$, with $\beta$ replaced by $\beta_{\rm eff}\approx3.5H_c$. Here $H_c$ is the Hubble rate at the transition, and $H=\sqrt\alpha\,H_c$ is the Hubble rate of the vacuum energy alone, with $\alpha=L/\rho_{\rm tot}$. 
 First collisions then account for all of the DM for a Dirac fermion and $\nu\ge2$ with $M_X$ between $5\times10^{12}$ and $2\times10^{17}$~GeV, if the inflationary Hubble rate lies below $2\times10^{10}$ to $2\times10^{11}$~GeV. 
 Strongly supercooled transitions, as in a classically scale-invariant $U(1)_{B-L}$ extension of the Standard Model~\cite{Ellis:2019oqb} or in hidden sectors~\cite{Katz:2016adq}, may also realize the benchmark if their walls collide before friction stops them, which remains to be checked.

\emph{A gravitational lower bound}.---For a runaway, vacuum-dominated transition with given $H/\beta$ and $\nu$, Eq.~\eqref{eq:Yuniv} times the expansion factor, about $e^{-4H/\beta}$, is a lower estimate. Every production process we left out adds to it. Near the highest masses, $\xi_\phi$ and loop corrections can lower it, by up to $20\%$ and $16\%$ times a logarithm, and contact terms and strong-gravity effects change it by about $1\%$ or less. 
 For other spins, the lower bound is Eq.~\eqref{eq:Yuniv} with their smallest $\chat$ over the unknown curvature coupling of the wall field, $1/6$ for a scalar and $13/6$ for a Proca field. 
After the first collision the walls continue as shells that cross many times~\cite{An:2026ghw}, and since $\sigma\propto\shat$, counter-streaming quanta above $M_X$ keep producing $X$. In the realization above this raises the yield by up to a factor of five if the hard quanta survive.
 We estimate that scatterings of hard quanta on the forming plasma there add, for $M_X\lesssim10^{15}$~GeV, up to $20\%$ for $\alpha=0.05$ and $80\%$ for $\alpha=5\times10^{-4}$.
 With the expansion correction, the curves of Fig.~\ref{fig:plane}, computed with $\xi_\phi=0$, are then upper bounds on $\Treh$ except near the upper mass edges, where the $\xi_\phi=1/6$ curve lies up to $8\%$ higher. Given a runaway transition, any stable particle with $M_X\lesssim\bar M_{\rm coll}$ that couples to the transition sector only through gravity is overproduced if $\Treh$ lies above the curve computed with $\xi_\phi=1/6$, which minimizes its yield. Like freeze-in~\cite{Garny:2015sjg}, this needs only gravity, but also a runaway transition with the given $H/\beta$ and $\nu$.

The lower bound assumes that the walls pass through each other at the first collision and that no entropy is produced afterward. A stage in which the field left by the collisions behaves like matter before it thermalizes would dilute the yield~\cite{Mansour:2026sdx}. It also requires a runaway transition. This holds for a transition in vacuum~\cite{An:2026ghw,Freese:2023fcr} and, in our reading of Refs.~\cite{Bodeker:2017cim,Gouttenoire:2021kjv}, for walls of a gauge singlet if the released vacuum energy far exceeds both the radiation energy and the leading-order friction pressure~\cite{Bodeker:2009qy,Caprini:2015zlo}. A runaway wall in a plasma carries only the part of the released energy not lost to this friction~\cite{Espinosa:2010hh}, and the yield falls with the square of that fraction. Walls that give mass to gauge bosons feel a friction that grows with $\gamma$~\cite{Bodeker:2017cim,Gouttenoire:2021kjv,Hoche:2020ysm} and run away only while $\gamma$ stays below a terminal value.
Particle production differs in the non-runaway regime~\cite{Abu-Ajamieh:2026jhn}.

\emph{Gravitational waves.}---The same bubble collisions source gravitational waves. After free passage, the walls continue as shells of hard quanta, which keep radiating until scatterings on the plasma degrade them~\cite{Strumia:2026uvz}. The spectrum therefore lies between the envelope limit, where the source ends at the collision~\cite{Caprini:2015zlo}, and the bulk-flow limit, where the shells persist~\cite{Jinno:2017fby,Konstandin:2017sat,Lewicki:2020jiv}. In both, the peak amplitude is set by $H/\beta$ and the peak frequency by $(\beta/H)\Treh$, about two times lower for bulk flow~\cite{Lewicki:2020jiv}. Measuring both would fix $H/\beta$ and $\Treh$, and the relic condition then $M_X$ on the light branch of the DM curve, where $R\simeq1$. Along the DM curves of Fig.~\ref{fig:plane}, the present peak frequency lies between $10^4$ and $10^7$~Hz in the envelope limit, above the bands of current and planned detectors. The strongest transitions of the realization above peak at $3\times10^4$ to $5\times10^5$~Hz in the bulk-flow limit, and the lower part of this range lies above the quantum limit of interferometric detection~\cite{Guo:2025cza}.
 A dedicated detector at these frequencies could probe the transition that produced the DM.

\emph{Conclusions.}---Runaway bubble collisions produce superheavy particles through gravity alone. The number of pairs depends only on the energies of the colliding walls. For exponential nucleation, the sum over first collisions is exact in flat space, and the mass reach is set by the duration of the transition and the Planck mass. For a given runaway transition, this gives a lower bound on the abundance of any stable particle that couples to the transition sector only through gravity. Its mass reach goes beyond that of freeze-in and, above the inflaton mass and $H_{\rm inf}$, of inflationary production, and it opens the region of heavy dark matter at low reheating temperature. The same bubble collisions emit gravitational waves above a few\,kHz, whose peak could constrain the transition and, through the relic condition, the DM mass.
The Supplemental Material gives the cross sections, the parton and classical-source derivations, the lattice tests, the first-collision sum, the dark-matter curves, the runaway conditions, the realization and the corrections beyond the first collision.
The code and the numerical results are available at \href{https://doi.org/10.5281/zenodo.24437869}{doi:10.5281/zenodo.24437869}.

\begin{acknowledgments}
This work was supported in part by the National Science and Technology Council (NSTC) Grant No.~NSTC-114-2112-M-002-020-MY3 (CWC), the Ministry of Education (Higher Education Sprout Project NTU-114L104022-1), the National Center for Theoretical Sciences of Taiwan (VQT), and the Vietnam National Foundation for Science and Technology Development (NAFOSTED) under grant number 103.01-2023.50 (VQT). We also acknowledge the use of AI-assisted tools for coding assistance, grammar and spelling checks, and language polishing of the manuscript.
\end{acknowledgments}

\bibliography{refs}

\clearpage
\onecolumngrid
\setcounter{secnumdepth}{1}
\setcounter{section}{0}
\setcounter{equation}{0}
\setcounter{figure}{0}
\setcounter{table}{0}
\renewcommand{\thesection}{S\arabic{section}}
\renewcommand{\theequation}{S\arabic{equation}}
\renewcommand{\thefigure}{S\arabic{figure}}
\renewcommand{\thetable}{S\Roman{table}}
\renewcommand{\theHsection}{S\arabic{section}}
\renewcommand{\theHequation}{S\arabic{equation}}
\renewcommand{\theHfigure}{S\arabic{figure}}
\renewcommand{\theHtable}{S\Roman{table}}

\begin{center}
{\large\bfseries Supplemental Material}
\end{center}
This Supplemental Material collects the derivations and checks behind the Letter. We use natural units, the metric $(+,-,-,-)$ and the reduced Planck mass $\MP=(8\pi G)^{-1/2}\simeq2.4\times10^{18}$~GeV.

\section{Graviton-mediated cross sections}\label{sec:S-xsec}

A quantum of the wall field $\phi$ from one wall and one from the other annihilate into a virtual graviton, which produces a pair of the stable particle $X$, of mass $M_X$. At the energies of interest, the $\phi$ quanta are effectively massless. Both sectors couple to the canonically normalized graviton $h_{\mu\nu}$, defined by $g_{\mu\nu}=\eta_{\mu\nu}+2h_{\mu\nu}/\MP$, through their stress tensors,
\begin{equation}\label{eq:S-Lint}
\mathcal L=-\frac{1}{\MP}\,h_{\mu\nu}\left(T^{\mu\nu}_\phi+T^{\mu\nu}_X\right).
\end{equation}
The scalar stress tensors include the curvature couplings $\xip$ of the wall field and $\xi_X$ of a scalar $X$, with $\mathcal L\supset-\xi\mathcal R\phi^2/2$, where $\mathcal R$ is the Ricci scalar. The amplitude is
\begin{equation}\label{eq:S-amp}
\mathcal M=-\frac{1}{\MP^2\shat}\Big(\mathcal T_\phi^{\mu\nu}\mathcal T_{X\,\mu\nu}-\frac12\mathcal T_\phi\mathcal T_X\Big),
\end{equation}
where $\mathcal T^{\mu\nu}$ are the matrix elements of the stress tensors between the vacuum and the two-particle states, $\mathcal T$ their traces, and $\shat$ the squared center-of-mass energy. Both matrix elements are conserved, so the gauge-dependent part of the propagator drops out.

The coupling $1/\MP$ has mass dimension $-1$, so the cross section grows with energy, $\sigma=c_X(\varrho)\,\shat/\MP^4$, with $\varrho=M_X^2/\shat$ and $v_X=\sqrt{1-4\varrho}$. A conserved $\mathcal T^{\mu\nu}$ splits into a traceless part and a trace, and the propagator contraction separates into the exchange of a spin-2 and of a spin-0 state. These do not interfere after the angular integration, so
\begin{equation}\label{eq:S-cTS}
960\pi\,c_X(\varrho)=\hat c_{\rm T}(\varrho)+(1-6\xip)^2\,\hat c_{\rm S}(\varrho),
\end{equation}
with the spin-2 part $\hat c_{\rm T}$ and the spin-0 part $\hat c_{\rm S}$ given by
\begin{equation}\label{eq:S-cTSforms}
\begin{aligned}
\hat c_{\rm T}^{\rm scalar}&=\frac16v_X^5, & \hat c_{\rm S}^{\rm scalar}&=\frac56v_X(1-6\xi_X+2\varrho)^2,\\
\hat c_{\rm T}^{\rm Dirac}&=v_X^3\Big(1+\frac83\varrho\Big), & \hat c_{\rm S}^{\rm Dirac}&=\frac{10}3v_X^3\varrho,\\
\hat c_{\rm T}^{\rm Proca}&=\frac{v_X}6(13+56\varrho+48\varrho^2), & \hat c_{\rm S}^{\rm Proca}&=\frac{5v_X}6(1-4\varrho+12\varrho^2).
\end{aligned}
\end{equation}
For the Dirac fermion $c_X$ counts $X\bar X$ pairs. A symmetry factor $1/2$ is included for identical scalars and vectors. Table~\ref{tab:S-c0} lists the massless coefficients $\chat\equiv960\pi\,c_X(0)$. Since $\xip$ is not known, the lower bound uses the smallest value over $\xip$, which is $1/6$ for a scalar, $1$ for the massless Dirac fermion and $13/6$ for a Proca field. 

\begin{figure}[t]
\centering
\includegraphics[width=0.9\textwidth]{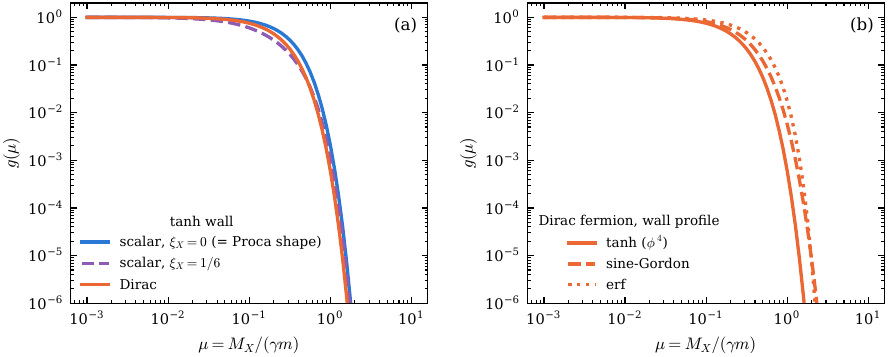}
\caption{Mass-suppression function $g(\mu)$, with $\mu=M_X/(\gamma m)$, for two colliding planar walls ($\xip=0$). (a) Minimally coupled scalar, conformally coupled scalar and Dirac fermion for the $\tanh$ profile. (b) Dirac fermion for the $\tanh$, sine-Gordon and error-function profiles.}
\label{fig:S-univ}
\end{figure}

The curvature couplings enter only the traces, and hence only the spin-0 part. The trace of the stress tensor of a massless Dirac fermion vanishes, so its massless coefficient is $\chat=1$ for every $\xip$. At finite mass $\xip$ enters the Dirac cross section through the term $\frac{10}{3}(1-6\xip)^2\varrho\ge0$. Folded with the wall spectra, its relative size scales as $\mu^2\ln^2(1/\mu)$, with $\mu=M_X/(\gamma m)$, where $\gamma$ is the Lorentz factor of the walls and $1/m$ their thickness (Sec.~\ref{sec:S-parton}). It therefore matters only within about a factor of ten of the highest masses the walls produce. This term is smallest at $\xip=1/6$. Figures~\ref{fig:S-univ}, \ref{fig:S-dmbench} and \ref{fig:S-anli} and Table~\ref{tab:S-window} use $\xip=0$, while the lower bound near the highest masses is the $\xip=1/6$ yield. It is lower by up to $20\%$ at the upper edge of the core, and by less than $10\%$ for $M_X\lesssim0.1\,\bar M_{\rm coll}$. 

\begin{table}[t]
\centering
\begin{tabular}{lcccc}
\hline
$X$ & $\chat(\xip)$ & $\xip=0$ & $\xip=1/6$ & threshold\\
\hline
scalar, $\xi_X=0$ & $\tfrac16+\tfrac56(1-6\xip)^2$ & $1$ & $\tfrac16$ & $v_X$\\
scalar, $\xi_X=\tfrac16$ & $\tfrac16$ & $\tfrac16$ & $\tfrac16$ & $v_X$\\
Dirac & $1$ & $1$ & $1$ & $v_X^3$\\
Proca & $\tfrac{13}6+\tfrac56(1-6\xip)^2$ & $3$ & $\tfrac{13}6$ & $v_X$\\
\hline
\end{tabular}
\caption{Massless coefficient $\chat=960\pi\,c_X(0)$ for each spin of $X$, as a function of the curvature coupling $\xip$ of the wall field and at $\xip=0$ and $1/6$. The last column gives the threshold behavior at $\xip=0$.}
\label{tab:S-c0}
\end{table}

\section{One collision of two planar walls}\label{sec:S-parton}

\emph{Parton picture.} In its rest frame a wall is a static profile $\phi_0(z)$ with field excursion $\Dphi$ and thickness of order $1/m$, which for the kink profiles is set by the mass $m$ of $\phi$ in the vacua. Its spectrum is the Fourier transform of its gradient, $F(k)=\int dz\,\phi_0'(z)e^{ikz}$, with $F(0)=\Dphi$ and $F$ exponentially small for $k\gg m$. A wall moving with Lorentz factor $\gamma\gg1$ depends on a single light-cone coordinate and carries the energy per unit wall area $A$, $E/A=\gamma\sigma_{\rm wall}$, with $\sigma_{\rm wall}=\int dz\,\phi_0'^2$ the tension. It is a superposition of massless plane waves, and in the quantum theory a coherent state of massless $\phi$ quanta, which we call partons. Their number per unit area and unit momentum is~\cite{Ghoshal:2026pew}
\begin{equation}\label{eq:S-dN}
\frac{dN}{A\,dp}=\frac{|F(p/\gamma)|^2}{\pi p}.
\end{equation}
A parton of momentum $p_1$ from one wall and one of momentum $p_2$ from the other collide head-on with $\shat=4p_1p_2$. The mean number of production events per unit area is
\begin{equation}\label{eq:S-NA}
\frac{N_{\rm ev}}{A}=\int_0^\infty dp_1\,dp_2\,\frac{dN_1}{A\,dp_1}\frac{dN_2}{A\,dp_2}\,\sigma(4p_1p_2)
=\int d\shat\,\frac{d\mathscr L}{d\shat}\,\sigma(\shat),
\end{equation}
where $d\mathscr L/d\shat$ is the pair luminosity per unit area. For two identical walls $\shat\,d\mathscr L/d\shat=(\Dphi^4/\pi^2)K(\zeta)$, with $\zeta=\shat/(4\gamma^2m^2)$, $K(\zeta)=\int_0^\infty(d\hat p/\hat p)\,W(\hat p)W(\zeta/\hat p)$ and $W(\hat p)=|F(m\hat p)|^2/\Dphi^2$. For the $\tanh$ profile this is Eq.~(3.19) in Ref.~\cite{Ghoshal:2026pew}. We use the exact $K$ throughout.

\emph{Energy factorization.} For $M_X\ll\gamma m$ the cross section is that of a massless $X$, $\sigma\simeq4c_X(0)p_1p_2/\MP^4$. It is proportional to the product of the two parton momenta, so Eq.~\eqref{eq:S-NA} factorizes. Each wall contributes $\int_0^\infty dp\,p\,dN_i/(A\,dp)=E_i/A$. For general $M_X$ we write
\begin{equation}\label{eq:S-theorem}
\frac{N_{\rm ev}}{A}=\frac{4c_X(0)}{\MP^4}\,\frac{E_1}{A}\,\frac{E_2}{A}\,g(\mu),\qquad
g(\mu)=\frac{\int_0^\infty d\zeta\,K(\zeta)\,c_X\big(\mu^2/(4\zeta)\big)}{c_X(0)\big[\int_0^\infty d\hat p\,W(\hat p)\big]^2},
\end{equation}
with $c_X(\varrho)=0$ for $\varrho>1/4$. Since $\int_0^\infty d\zeta\,K(\zeta)=[\int_0^\infty d\hat p\,W(\hat p)]^2$, the mass-suppression function obeys $g(0)=1$ exactly, for any wall profile. The parton density grows as $1/p$ at small $p$, but the cross section weights each parton by its energy, so soft partons contribute little. Partons with $p<0.1\,\gamma m$ carry $11\%$ (error function) to $19\%$ ($\tanh$) of $E_i/A$. The product $E_1E_2$ is invariant under boosts along the collision axis.

\emph{Classical source.} We derive the same result by treating the colliding walls as a classical gravitational source. Ultra-relativistic walls pass through each other with superposed profiles, up to corrections of order $1/\gamma^2$~\cite{Strumia:2026uvz}. During this free passage $\phi=\phi_1(t+x)+\phi_2(t-x)$. A single uniformly moving wall cannot produce particles~\cite{Zeldovich:1977vgo,Frieman:1985fr,Gibbons:1975jb}. The cross term of the stress tensor,
\begin{equation}\label{eq:S-Tcross}
T^{\mu\nu}_\times=\phi_1'\,\phi_2'\,\big[n^\mu\bar n^\nu+\bar n^\mu n^\nu-2\eta^{\mu\nu}\big]
+2\xip\,(\eta^{\mu\nu}\Box-\partial^\mu\partial^\nu)\big[\phi_1\phi_2\big],
\end{equation}
with $n=(1,1,0,0)$ and $\bar n=(1,-1,0,0)$, has a Fourier transform proportional to $F_1(p_1/\gamma)F_2(p_2/\gamma)$ times the two-parton matrix element of Eq.~\eqref{eq:S-amp}. Squaring and integrating over the two-body phase space of $X$ reproduces Eq.~\eqref{eq:S-NA} exactly at $\mathcal O(\MP^{-4})$, with no additional coherence factor. For a massless, conformally coupled $X$, the master formula, Eqs.~(2.2)--(2.5) in Ref.~\cite{Redi:2026nzi}, applied to $T^{\mu\nu}_\times$, gives the same result. The cross term of the potential lacks the factor $\gamma^2$ of the gradient terms and matters only for partons with momenta of a few $m$.

\emph{Mass suppression and profiles.} At finite mass $g(\mu)$ depends on the spin and on the wall profile (Fig.~\ref{fig:S-univ}). We use the $\tanh$, sine-Gordon and error-function profiles, with tensions $m\Dphi^2/6$, $2m\Dphi^2/\pi^2$ and $m\Dphi^2/(2\sqrt\pi)$. For the Dirac fermion $g$ falls to $0.1$ at $\mu=0.42$, $0.53$ and $0.68$, as shown in Fig.~\ref{fig:S-univ}(b). Compared at fixed wall energy and field excursion, which is the comparison relevant in cosmology, the three $10\%$ points agree within $10\%$, and below them the yields differ by at most a factor of $1.3$. We use the $\tanh$ profile throughout. 

\begin{figure}[t]
\centering
\includegraphics[width=0.9\textwidth]{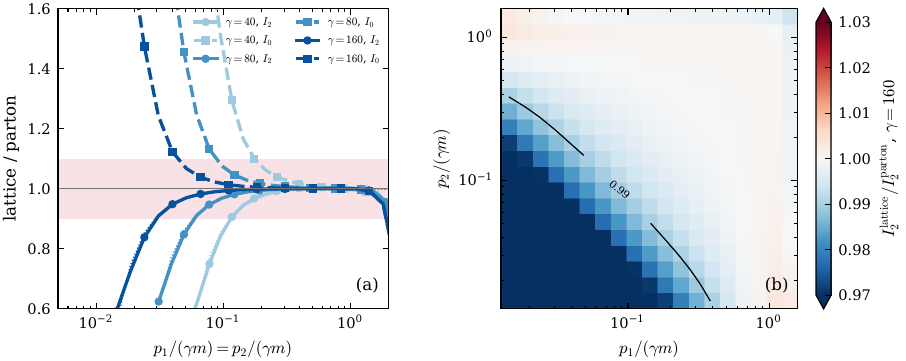}
\caption{Ratio of lattice to parton results for the Fourier-transformed stress tensor in a kink--antikink collision in $1+1$ dimensions, extrapolated to zero lattice spacing. (a) Along $p_1=p_2$, for $\gamma=40$, $80$ and $160$. Solid with circles: the spin-2 invariant $I_2$. Dashed with squares: the trace invariant $I_0$. Dotted: $I_2$ for the two walls superposed without interaction. The curves start at $p_1=p_2=2m$. Band: $\pm10\%$. (b) $I_2$ over the $(p_1,p_2)$ plane at $\gamma=160$, for $p_1,p_2$ between $2m$ and $1.5\gamma m$. Contour: $0.99$.}
\label{fig:S-lattice}
\end{figure}

\begin{figure}[t]
\centering
\includegraphics[width=0.5\textwidth]{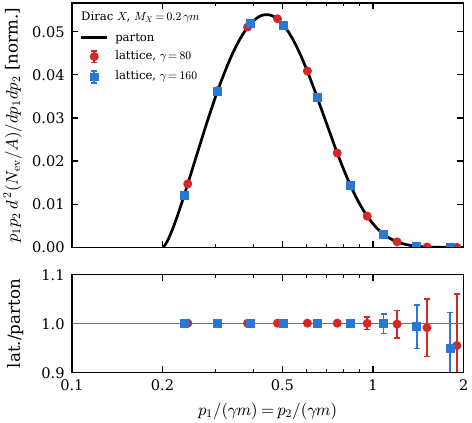}
\caption{Differential yield of pairs of a Dirac $X$ with $M_X=0.2\gamma m$ per unit wall area along $p_1=p_2$, multiplied by $p_1p_2$ and in units of $4c_X(0)\Dphi^4\gamma^2m^2/(\pi^2\MP^4)$. Curve: parton prediction. Points: lattice simulation of a kink--antikink collision in $1+1$ dimensions at $\gamma=80$ (circles) and $160$ (squares), extrapolated to zero lattice spacing. Lower panel: ratio of lattice to parton. Error bars: numerical uncertainty of the lattice.}
\label{fig:S-lattice-yield}
\end{figure}

\section{Lattice test of free passage}\label{sec:S-lattice}

The parton picture assumes that the walls pass through each other with superposed profiles. We test this with nonlinear simulations in $1+1$ dimensions of a kink colliding with an antikink in the potential $\frac{\lambda}{4}(\phi^2-v_{\rm s}^2)^2$, with quartic coupling $\lambda$, minima at $\phi=\pm v_{\rm s}$ and vacuum mass $m=\sqrt{2\lambda}\,v_{\rm s}$, at $\gamma=10$, $40$, $80$ and $160$. After integration over the final-state phase space, the production rate depends on the source only through
\begin{equation}
I_2=\tilde T^*_{\mu\nu}\tilde T^{\mu\nu}-\frac13|\tilde T^\mu{}_\mu|^2,\qquad I_0=|\tilde T^\mu{}_\mu|^2,
\end{equation}
the traceless (spin-2) and the trace (spin-0) parts of the Fourier-transformed stress tensor $\tilde T^{\mu\nu}$. A massless $X$ feels only $I_2$, and a massive $X$ also feels $I_0$. For free passage, $I_0=6I_2$ at every pair of parton momenta.

\emph{Method and accuracy.} We evolve the field on a lattice, accumulate $\tilde T^{\mu\nu}$ in a Gaussian time window around the collision, and compare with the parton prediction on a grid of $28\times28$ parton momenta $p_1,p_2$ between $m/2$ and $3\gamma m$. Each run uses $20$ and $40$ lattice points per wall thickness, and we extrapolate to zero spacing. The error falls by about four between the two runs, as expected for second-order time stepping, and we take the size of the extrapolation as the uncertainty. It is at most $0.3\%$ for $p_1,p_2$ between $0.1\gamma m$ and $0.7\gamma m$ and grows to $1$ to $2\%$ at $\gamma m$. A wider window and a smaller time step change the results there by less than $10^{-4}$. We also pass the superposition $\phi_1+\phi_2$ of the two uniformly moving walls through the same steps. Without its potential energy, it reproduces the parton prediction to $5\times10^{-5}$ at all momenta, which checks the whole procedure.

\emph{Invariants.} Figure~\ref{fig:S-lattice}(a) shows $I_2$ and $I_0$ along $p_1=p_2$ for $\gamma=40$, $80$ and $160$. For $p_1,p_2$ between $0.1\gamma m$ and $\gamma m$, $I_2$ agrees with the parton prediction within $9.4\%$, $1.9\%$ and $0.6\%$ at these $\gamma$, and $I_0$ within $29\%$, $4.6\%$ and $1.3\%$. The departure from free passage itself, the ratio of lattice to superposition, is $1.3\%$, $0.3\%$ and $0.1\%$ for $I_2$. It falls roughly as $1/\gamma^2$, as expected in Ref.~\cite{Strumia:2026uvz}. For $I_0$ it is $0.5\%$ and $0.1\%$ at $\gamma=80$ and $160$. At softer momenta, both invariants deviate from the parton prediction. $I_2$ falls by about $2(m/p)^2$, and $I_0$ exceeds the prediction by up to a factor of two at a few $m$. The superposition with its potential energy shows the same deviations within $20\%$ at $\gamma=80$ and $160$. They come from the potential cross term $V(\phi_1+\phi_2)-V(\phi_1)-V(\phi_2)$, which the parton prediction drops. They sit at fixed $p/m$, so they move to smaller $p/(\gamma m)$ as $\gamma$ grows. Figure~\ref{fig:S-lattice}(b) shows the ratio for $I_2$ over the $(p_1,p_2)$ plane at $\gamma=160$. It stays within $0.7\%$ of one wherever $p_1p_2>(0.1\gamma m)^2$, and the deficit is confined to smaller $p_1p_2$. At $\gamma=10$ (not shown), the hard region is not well separated from $m$, and $I_2$ exceeds the prediction by $6$ to $8\%$ between $6m$ and $8m$.

\emph{Yield.} At given parton momenta, the ratio of lattice to parton yield is $f_2r_2+(1-f_2)r_0$, with $r_i=I_i^{\rm lattice}/I_i^{\rm parton}$ and $f_2=\hat c_{\rm T}/(\hat c_{\rm T}+\hat c_{\rm S})$, the spin-2 share of the cross section in Eq.~\eqref{eq:S-cTS} at $\xip=0$. For a Dirac $X$, $f_2$ falls from $1$ at $M_X\to0$ to $2/3$ at threshold. Figure~\ref{fig:S-lattice-yield} shows the differential yield along $p_1=p_2$ for $M_X=0.2\gamma m$, where both parts contribute, and Fig.~1 of the Letter shows it for $M_X=10^{-3}\gamma m$. For the lighter mass, the region $p_1=p_2>0.2\gamma m$ gives about $80\%$ of the yield along the diagonal. There, lattice and parton agree within $0.3\%$ up to $0.7\gamma m$ at $\gamma=80$ and $160$, and within the numerical uncertainty beyond. The uncertainty is $0.3\%$ at $0.2\gamma m$ and $1$ to $2\%$ at $\gamma m$. For $M_X=0.2\gamma m$, whose threshold lies at $p_1=p_2=0.2\gamma m$, they agree within $0.12\%$ up to $\gamma m$. Integrated over the grid, the lattice yield agrees with the parton yield within $0.8\%$ at $\gamma=80$ for $\mu\le1$, and at $\gamma=160$ within $0.2\%$ for $\mu\le0.5$ and $1.3\%$ at $\mu=1$. These are central values. The extrapolation to zero lattice spacing moves them by about $1\%$ at $\mu=0.5$ and by $3$ to $4\%$ at $\mu=1$, which we take as their uncertainty. At $\gamma=10$ it is larger by up to $13\%$ near $\mu\simeq0.3$.

\emph{Quantum scattering and related work.} The lattice is classical. A hard parton of momentum $p$ can also scatter on the partons of the other wall through the quartic coupling. With the parton density of Eq.~\eqref{eq:S-dN}, cut off at momenta of order $m$, its probability is about $0.03\,\lambda\,m/p$, an estimate, and gauge couplings of a Higgs wall give a similar suppression by $m/p$. For $p\gtrsim0.1\gamma m$ and $\gamma\gtrsim10^3$ this is far below the lattice uncertainty. Reference~\cite{An:2026sdu} shows that production through couplings linear in $\phi$ depends on the potential through $V'$ at the field value reached at the collision. The graviton couples to the stress tensor instead, whose potential part is the cross term discussed above. The order-one suppression from the finite bubble radius found there, relative to parallel walls of a single size, is a property of the bubble population, which the first-collision sum of Sec.~\ref{sec:S-geom} includes.

\emph{Other potentials.} Walls accelerated by a tilted quartic potential change their profile but still pass freely. With the spectra measured just before the collision, the parton cross term reproduces the lattice within $6\%$ ($I_2$) and $13\%$ ($I_0$) for walls that start at $\gamma_0=20$. For walls that start almost at rest, the test is inconclusive. Such walls develop an overshoot and an oscillating wake, and their measured spectrum has zeros that make the comparison ill-conditioned. 
 In an even sextic potential, as for the radial Higgs field, the walls are trapped and turn back after the first crossing. The first collision is not affected, and the parton cross term reproduces the lattice there within $4\%$.

\section{Oblique collisions and the first-collision sum}\label{sec:S-geom}

\emph{Oblique factor.} In three dimensions the two walls that meet at a point of the collision surface move along unit normals $\mathbf n_1$ and $\mathbf n_2$ at an angle $\theta$. We treat each wall near that point as a planar null sheet of partons and define $\eperp\equiv\frac12|\mathbf n_1-\mathbf n_2|=\sin(\theta/2)$, so that $\eperp=1$ is a head-on collision. Partons then collide with $\shat=4p_1p_2\eperp^2$, and the event rate per volume carries the M{\o}ller velocity $v_{\rm M}=1-\cos\theta=2\eperp^2$. The time integral of the two sheets $\delta(\mathbf n_i\cdot\mathbf x-t)$ leaves $\delta((\mathbf n_1-\mathbf n_2)\cdot\mathbf x)$, whose integral across the collision surface is $1/(2\eperp)$. A patch $dA_{\rm locus}$ of the collision surface therefore yields
\begin{equation}\label{eq:S-oblique}
\frac{dN_{\rm ev}}{dA_{\rm locus}}=\frac{N_1}{A}\frac{N_2}{A}\,\sigma(\shat)\,\eperp
=\frac{4c_X(0)}{\MP^4}\frac{E_1}{A}\frac{E_2}{A}\,\eperp^3\,g\Big(\frac{M_X}{M_{\rm coll}}\Big),
\end{equation}
with the collision energy scale $M_{\rm coll}=\sqrt{\gamma_1\gamma_2}\,m\eperp$. Here $N_i/A$ are the parton numbers per unit area at the given momenta, and $\gamma_{1,2}$ are the Lorentz factors of the two walls. The factor $\eperp^3$ combines the flux factor $\eperp$ with $\eperp^2$ from $\sigma\propto\shat$. We checked it three ways. A boost along $\mathbf n_1+\mathbf n_2$ makes the walls antiparallel and gives the same result. The geometric factor of Ref.~\cite{An:2026sdu}, per unit area of the collision surface, equals our flux factor. A direct integration over two crossing sheets of Gaussian thickness reproduces the flux factor within $1.2\%$ for $\theta$ between $30^\circ$ and $180^\circ$, except for a $4\%$ deficit at $90^\circ$ from the edges of the box, which halves in a box twice as large.

\emph{First collisions.} We take a runaway transition in vacuum with nucleation rate $\Gamma\propto e^{\beta t}$ and $\beta\gg H$, with $H$ the Hubble rate, so that space is flat during the transition. The walls carry all of the latent heat $L$. Energy conservation gives $E/A=Lr/3$ for a wall of radius $r$, the vacuum energy sets $H^2=L/(3\MP^2)$, and the released energy thermalizes into $g_*=106.75$ degrees of freedom at the reheating temperature $\Treh$, with $L=(\pi^2g_*/30)\Treh^4$. We count only first collisions, the surfaces where the walls of two bubbles arrive at the same time and before any other wall. These are the interfaces of a Johnson--Mehl tessellation, which we follow in a volume $V$. Writing $E_i/A=(L/\beta)\hat e_i$ with $\hat e_i=\beta r_i/3$, the number density of $X$ is
\begin{equation}\label{eq:S-nX}
n_X=\frac{8c_X(0)L^2Q_\beta}{\beta\MP^4}\,R(M_X),\qquad
Q_\beta=\frac{1}{\beta V}\int dA_{\rm locus}\,\hat e_1\hat e_2\eperp^3,
\end{equation}
where $R\le1$ is the average of $g$ over the surfaces, weighted by $\hat e_1\hat e_2\eperp^3$. Each event gives two particles.

\emph{Exact geometric factor.} For exponential nucleation the two radii and $\eperp$ are statistically independent on the surfaces, so $Q_\beta=(A_{\rm locus}/\beta V)\langle\hat e_1\rangle\langle\hat e_2\rangle\langle\eperp^3\rangle$ with area averages. Three facts give the factors. First, the interface area per volume is $A_{\rm locus}/V=2\beta/3$, from Eq.~(70) in Ref.~\cite{Miles:1972} for exponential nucleation~\cite{Horalek:1988}. Second, a surface element reached at time $t_c$ by a bubble of radius $r_i$ receives the weight $r_i^2\Gamma(t_c-r_i)\propto r_i^2e^{-\beta r_i}$, the area of the sphere of possible nucleation sites times the earlier nucleation rate. The Poisson void probability depends only on $t_c$~\cite{Moller:1992}. So $\beta r_i$ follow independent Gamma distributions of shape $3$, and $\langle\hat e_i\rangle=1$. Third, the interface between bubbles nucleated at $(t_1,\mathbf x_1)$ and $(t_2,\mathbf x_2)$ is the zero set of $f(\mathbf x)=|\mathbf x-\mathbf x_1|-|\mathbf x-\mathbf x_2|-(t_2-t_1)$, whose area per unit volume is $\int\delta(f)\,|\nabla f|$, with $|\nabla f|=|\mathbf n_1-\mathbf n_2|=2\eperp$. Isotropy makes $\cos\theta$ uniform, which gives $\eperp$ the density $2\eperp$, and the factor $|\nabla f|\propto\eperp$ turns it into $3\eperp^2$, so $\langle\eperp^3\rangle=1/2$. 
 Hence
\begin{equation}\label{eq:S-Qbeta}
Q_\beta=\frac23\cdot1\cdot1\cdot\frac12=\frac13 .
\end{equation}
Oblique collisions halve the yield. Assuming head-on collisions everywhere would give $2/3$. A Monte Carlo of the first-collision surfaces in a periodic box, with a control variate for the super-Poissonian fluctuations from rare early bubbles, gives $Q_\beta=0.336\pm0.002$ from eight independent realizations in a box of side $32/\beta$. An independent estimator based on line crossings gives $0.3326\pm0.0023$. A simpler estimate~\cite{Ghoshal:2026pew} takes all bubbles as spheres of a single radius $R_*=(8\pi)^{1/3}/\beta$, with $E/A=LR_*/3$, $n_X=(3/R_*)N_{\rm ev}/A$ and head-on collisions. It exceeds our yield by $(8\pi)^{1/3}/(6Q_\beta)\approx1.5$. Oblique collisions halve the yield, and the spread of bubble radii raises it by $4/(8\pi)^{1/3}\approx1.37$ relative to a single radius.

\emph{Yield and collision scale.} With $c_X(0)=\chat/(960\pi)$ and the entropy density $s=(2\pi^2g_*/45)\Treh^3$, Eq.~\eqref{eq:S-nX} gives
\begin{equation}\label{eq:S-Yuniv}
Y_X=\frac{n_X}{s}=\frac{\chat}{480}\sqrt{\frac{g_*}{10}}\,\frac{H}{\beta}\Big(\frac{\Treh}{\MP}\Big)^3R(M_X),
\end{equation}
with $\sqrt{g_*/10}/480\simeq6.8\times10^{-3}$. For first collisions and free passage in flat space the prefactor is exact. Weighted by the event yield, the radii follow Gamma distributions of shape $4$ and $\eperp$ has the density $6\eperp^5$. With $\sigma_{\rm wall}\gamma_i=Lr_i/3$ and $\sigma_{\rm wall}=m\Dphi^2/6$, the yield-weighted collision scale is
\begin{equation}\label{eq:S-Mbar}
\bar M_{\rm coll}=\frac{525\pi}{256}\,\frac{L}{\beta\Dphi^2}\simeq11\,\frac{H}{\beta}\,\frac{\MP}{\nu^2},
\end{equation}
where $\Dphi=\nu\Lq$. The yield-weighted average of $g$ over the collisions, $R(M_X)$, is $0.9$, $0.5$ and $0.1$ at $M_X/\bar M_{\rm coll}=0.04$, $0.17$ and $0.45$. For $\nu=3$, $\bar M_{\rm coll}$ is $0.12\,\MP$ at $H/\beta=0.1$ and $0.012\,\MP$ at $H/\beta=0.01$, independent of $\Treh$. Most of the yield comes from bubbles of radius about $4\beta^{-1}$, so the expansion of the universe during the transition corrects Eq.~\eqref{eq:S-Yuniv} at relative order $4H/\beta$. Equation~\eqref{eq:S-Yuniv} counts pairs from the thin walls and applies for $M_X\gg\beta$.

\emph{Expansion during the transition.} We repeated the first-collision sum on the expanding background. The rate per physical volume is $\Gamma=\beta^4e^{\beta t}/(8\pi)$, the false vacuum dominates with $H^2=L/(3\MP^2)$, the released energy redshifts as radiation, and the walls carry the vacuum energy released inside their bubbles. The false-vacuum fraction and the scale factor are solved together. The first-collision surfaces are found in comoving coordinates, and the entropy follows from the late radiation. At fixed latent heat and fixed $\beta$ the yield relative to Eq.~\eqref{eq:S-Yuniv} is $0.669\pm0.006$, $0.857\pm0.012$ and $0.951\pm0.011$ for $H/\beta=0.1$, $0.03$ and $0.01$, from boxes of side $24/\beta$ and $32/\beta$. These values follow $e^{-4.2H/\beta}$ within about $3\%$. The ratio tends to one as $H/\beta\to0$, and a finer surface resolution or grid changes it within the errors. The expansion lowers the yield because the walls lose energy to redshift and the released energy is diluted while the transition is still going on. The dark-matter curves, computed with Eq.~\eqref{eq:S-Yuniv}, move up in $\Treh$ by at most $(0.67)^{-1/3}\simeq1.14$.

\emph{Comparison with the soft part of the source.} Reference~\cite{Redi:2026nzi} estimates the production by the part of the source that varies on the time scale $1/\beta$, for light particles, as $Y\sim5\times10^{-4}c_{\rm DM}(\beta/H)(\Treh/\MP)^3\,\Omega_{\rm GW}/\Omega_\gamma$ [Eq.~(3.18) in Ref.~\cite{Redi:2026nzi}, with its $T_*$ and $H_*$ set to $\Treh$ and $H$], with $\Omega_\gamma=5\times10^{-5}$ and $c_{\rm DM}=8$ for a Dirac fermion in their normalization. With the present gravitational wave (GW) density of the envelope approximation, $\Omega_{\rm GW}\simeq2.9\times10^{-6}(H/\beta)^2$, this gives $Y\simeq2.3\times10^{-4}(H/\beta)(\Treh/\MP)^3$. Both scale as $H/\beta$, and the hard part, Eq.~\eqref{eq:S-Yuniv}, is about thirty times larger for every $H/\beta$. Since Eq.~(3.18) in Ref.~\cite{Redi:2026nzi} is an order-of-magnitude estimate, so is this ratio.

\section{Thermal freeze-in and the dark-matter curves}\label{sec:S-fi}

\emph{Freeze-in.} Graviton exchange also produces $X$ in the thermal Standard Model plasma. For instantaneous reheating to $T_{\max}=\Treh$, massless Standard Model particles with Maxwell--Boltzmann statistics, and the formalism of Ref.~\cite{Redi:2026nzi},
\begin{equation}\label{eq:S-YFI}
Y_{\rm FI}=4.8\times10^{-5}\,\hat c_{\rm FI}\Big(\frac{\Treh}{\MP}\Big)^3\mathcal F_X\Big(\frac{M_X}{\Treh}\Big),
\end{equation}
where $\mathcal F_X(0)=1$ and $\mathcal F_X$ falls exponentially for $M_X\gtrsim\Treh$. In the plasma, only the Higgs scalars source the spin-0 part, so $\hat c_{\rm FI}=\hat c_{\rm T}+(4/283)(1-6\xi_H)^2\hat c_{\rm S}$. Here $\xi_H$ is the curvature coupling of the Higgs field, and $4$ is the spin-2 weight of the Higgs scalars in units of one real scalar. The weight of the whole Standard Model plasma is $283=4+3\times45+12\times12$, because the central charges of a scalar, a Weyl fermion and a gauge field are in the ratio $1:3:12$~\cite{Redi:2026nzi}. For a massless Dirac fermion, $\hat c_{\rm FI}=1$. 
 For $M_X\ll\Treh$,
\begin{equation}\label{eq:S-ratioFI}
\frac{Y_X}{Y_{\rm FI}}=142\,\frac{\chat}{\hat c_{\rm FI}}\,\frac{H}{\beta},
\end{equation}
so for a Dirac fermion collisions dominate once $H/\beta\gtrsim7\times10^{-3}$. Above $\Treh$ freeze-in is Boltzmann-suppressed, while the collision yield stays flat up to $\bar M_{\rm coll}$.

\begin{figure}[t]
\centering
\includegraphics[width=0.9\textwidth]{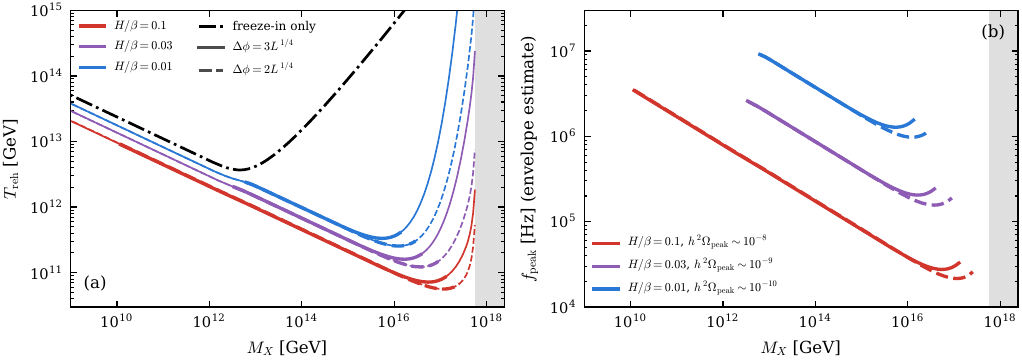}
\caption{(a) Reheating temperature that gives $\Omega_Xh^2=0.12$ for a gravity-only Dirac fermion, from first collisions with $\tanh$ walls plus thermal freeze-in, for $H/\beta=0.1$, $0.03$ and $0.01$ (red, purple, blue) and $\nu=3$ (solid) and $2$ (dashed). Thick segments mark the core. Dash-dotted curves show freeze-in alone. The shaded region is $M_X>\Lambda_{\rm cut}/2$. (b) Present-day GW peak frequency along the cores in the envelope approximation~\cite{Caprini:2015zlo}.}
\label{fig:S-dmbench}
\end{figure}

\emph{Dark-matter curves.} The observed dark-matter (DM) abundance $\Omega_Xh^2=0.12$~\cite{Planck:2018vyg},
\begin{equation}\label{eq:S-relic}
M_X\big[Y_X+Y_{\rm FI}\big]=4.37\times10^{-10}\,{\rm GeV},
\end{equation}
fixes $\Treh$ as a function of $M_X$ for given $H/\beta$ and $\nu$ (Fig.~\ref{fig:S-dmbench}). We keep the core of each curve, defined by four conditions. (i) $R\ge0.1$, so that the yield does not rest on the profile-dependent tail of the wall spectrum. (ii) Freeze-in is below $10\%$ of the collision yield. (iii) Parton collision energies lie below the gravitational cutoff $\Lambda_{\rm cut}=0.46\,\MP$, and the yield above it is discarded. (iv) $M_X\ge10\beta$. Where $R\simeq1$ the curve falls as $\Treh\propto(M_XH/\beta)^{-1/3}$, and near $\bar M_{\rm coll}$ it turns up. The cores span $M_X$ from $10^{10}$ to $3\times10^{17}$~GeV, with $\Treh$ from a few $\times10^{10}$ to $10^{13}$~GeV. The lowest reheating temperature on a core is $5.6\times10^{10}$~GeV.

\emph{Gravitational cutoff.} Since $\sigma\propto\shat$, the yield is dominated by the most energetic parton pairs. We take the cutoff at the species scale $M_{\rm Pl}/\sqrt N$~\cite{Dvali:2007hz}, with $M_{\rm Pl}=\sqrt{8\pi}\,\MP$ and $N\approx118$ Standard Model degrees of freedom ($28$ bosonic and $90$ fermionic, without the factor $7/8$ of $g_*$), which gives $0.46\,\MP$. The argument does not fix the $\mathcal O(1)$ factor. With $\MP$ in place of $M_{\rm Pl}$ it gives $0.09\,\MP$. Tree-level unitarity of $s$-channel graviton exchange fails only at $\sqrt{\shat}\approx2.3\,\MP$~\cite{Han:2004wt}. The hard cut is a test of sensitivity, not a bound. For $H/\beta=0.1$, $\bar M_{\rm coll}$ lies within a factor of four of $\Lambda_{\rm cut}$, so the heaviest masses depend on the ultraviolet completion of gravity.

\emph{Tensor-mode bound and freeze-in overproduction.} The absence of tensor modes in the cosmic microwave background (CMB), a tensor-to-scalar ratio $r_{0.05}<0.036$~\cite{BICEP:2021xfz}, with the scalar amplitude $A_s=2.1\times10^{-9}$~\cite{Planck:2018vyg}, bounds the inflationary Hubble rate,
\begin{equation}\label{eq:S-Hinf}
H_{\rm inf}<\pi\MP\sqrt{A_sr_{0.05}/2}=4.7\times10^{13}\,{\rm GeV}.
\end{equation}
For instantaneous reheating, $3H_{\rm inf}^2\MP^2=(\pi^2g_*/30)\Treh^4$, this bounds $\Treh<5.8\times10^{15}$~GeV. The bound applies to both channels, because the vacuum energy of the transition cannot exceed the inflationary one. Above the freeze-in curve of Eq.~\eqref{eq:S-relic} with $Y_X=0$, freeze-in alone gives $\Omega_Xh^2>0.12$ for any stable particle that couples only through gravity. Instantaneous reheating gives the smallest freeze-in for $M_X\gtrsim\Treh$. With slower reheating the plasma reaches $T_{\max}>\Treh$, and production near $T_{\max}$ escapes much of the Boltzmann suppression [Eq.~(23) in Ref.~\cite{Garny:2017kha}]. The overproduction region is therefore conservative there. For $M_X\ll\Treh$ slower reheating changes freeze-in by a factor of order one, which we have not computed. Below the CMB bound, freeze-in DM requires $\Treh\ge3.7\times10^{12}$~GeV and has $M_X/\Treh\le21$, so its heaviest mass is about $1.2\times10^{17}$~GeV, reached only near the bound. With slower reheating the maximal temperature exceeds $\Treh$ but stays of order the instantaneous value for the same $H_{\rm inf}$ [Eq.~(13) in Ref.~\cite{Garny:2017kha}], so this cap remains. Along the collision cores $M_X/\Treh$ reaches $3\times10^4$ to $4\times10^6$.

\emph{Production at the end of inflation.} Gravity also produces $X$ at the end of inflation. For $M_X\lesssim H_{\rm inf}$ and reheating after $H$ falls to $M_X$, Ref.~\cite{Ema:2019yrd} estimates for a Majorana fermion $\rho_X/s\simeq\mathcal A\,M_X^2T_R/(4\MP^2)$, where $\rho_X$ is the energy density of $X$, $\mathcal A\sim10^{-3}$ and $T_R$ the reheating temperature. A Dirac fermion gives twice this. Above $H_{\rm inf}$ the coefficient $\mathcal A$ is replaced by $\mathcal A(M_X/H_{\rm inf})^2e^{-c_{\rm inf}M_X/H_{\rm inf}}$ with $c_{\rm inf}$ of order one. With $T_R=\Treh$ and $c_{\rm inf}=1$, keeping this contribution below $10\%$ of the DM along the cores requires $M_X/H_{\rm inf}\gtrsim13$ to $44$, for example $H_{\rm inf}\lesssim4\times10^{11}$~GeV at $M_X=10^{13}$~GeV and $\Treh=10^{12}$~GeV. Above $M_X\approx2\times10^{15}$~GeV the CMB bound $H_{\rm inf}<4.7\times10^{13}$~GeV alone satisfies the condition on every curve, so the heavy end of the window needs no assumption on inflation. Below that mass the DM curves assume this condition. The required ratio scales as $1/c_{\rm inf}$  and depends on how inflation ends. For an abrupt end, our Bogoliubov calculation of Sec.~\ref{sec:S-anli} needs $70$ to $180$, which would move the threshold to $(3$--$8)\times10^{15}$~GeV. If the transition itself dominates after an earlier reheating, its entropy dilutes these relics and relaxes the condition.

\emph{Black holes and gravitational waves.} For $0.01\le H/\beta\le0.1$ the probability that a Hubble patch collapses into a black hole, extrapolated from Eq.~(15) in Ref.~\cite{Gouttenoire:2023naa}, is below $10^{-40}$. For walls that pass through each other the GW spectrum lies between two limits~\cite{Strumia:2026uvz}. In the envelope limit the source ends at the collision. For a vacuum transition~\cite{Caprini:2015zlo} the peak then has $h^2\Omega_{\rm peak}\sim10^{-6}(H/\beta)^2$ and a present-day frequency between $10^4$ and $10^7$~Hz along the cores. In the bulk-flow limit the walls continue as shells that conserve their energy~\cite{Jinno:2017fby,Konstandin:2017sat}. Thin-wall simulations then give the peak at the angular frequency $0.56\beta$ instead of $1.16\beta$, a peak amplitude within $10\%$ of the envelope one, and slopes $\omega^{1.05}$ and $\omega^{-2.10}$ instead of $\omega^{3.03}$ and $\omega^{-0.88}$ [Table~I in Ref.~\cite{Lewicki:2020jiv}]. After the collision the walls become pulses of hard quanta, which stop radiating once scatterings on the plasma have degraded them, after a time $t_{\rm stop}$~\cite{Strumia:2026uvz}. For hard $2\to2$ scatterings, Eq.~(28) in Ref.~\cite{Strumia:2026uvz} gives $t_{\rm stop}/R_c\approx1.2\,E_c\beta/(k\MP H)$, where $R_c$ is the bubble radius at the collision, $E_c$ the energy of a wall quantum and $k\lesssim1$ a coupling factor. With $E_c$ of order $\bar M_{\rm coll}$ this is about $13/(k\nu^2)$, an estimate, so the shells persist for one to a few bubble radii and the spectrum lies near the bulk-flow limit around the peak. Medium-induced splittings into gauge-charged particles can stop the shells within $R_c$, and the spectrum then approaches the envelope limit. The peak frequency is therefore uncertain by a factor of about two.

\section{When do walls run away?}\label{sec:S-runaway}

The yield assumes that the walls carry the released vacuum energy up to their first collision. This section summarizes known results. We do not derive them.

In a transition in vacuum, no plasma resists the walls, and they run away. In a plasma in equilibrium ahead of the wall at temperature $T$, particles whose mass changes by $\Delta m_a$ across the wall exert a pressure that tends, for $\gamma\to\infty$, to~\cite{Bodeker:2009qy}
\begin{equation}
P_{\rm LO}=\sum_ac_a\frac{\Delta m_a^2T^2}{24},
\end{equation}
with $c_a=1$ ($1/2$) per bosonic (fermionic) degree of freedom. At this order the walls run away if the released vacuum energy exceeds $P_{\rm LO}$, that is, if the ratio $\alpha_{\rm r}$ of the released vacuum energy $L$ to the radiation energy density exceeds $\alpha_\infty\simeq30\sum_ac_a\Delta m_a^2/(24\pi^2g_*T^2)$~\cite{Caprini:2015zlo}.

At next-to-leading order, gauge bosons whose mass changes across the wall are emitted as transition radiation, and the pressure grows with the Lorentz factor. Refs.~\cite{Bodeker:2017cim,Gouttenoire:2021kjv} find $P_{1\to2}\sim\gamma g_V^2m_VT^3$, with $g_V$ the gauge coupling and $m_V$ the vector-boson mass, which limits the walls to $\gamma_{\rm term}\sim(L-P_{\rm LO})/(g_V^2m_VT^3)$. Reference~\cite{Hoche:2020ysm} resums multiple soft emissions and finds a pressure that grows as $\gamma^2T^4$. The scaling with $\gamma$ is therefore debated. In both cases, walls charged under a gauge group reach a terminal Lorentz factor. They then carry only part of the released energy~\cite{Ellis:2019oqb}, which lowers both the yield and the mass reach. Such walls behave as runaway walls up to their first collision only if their Lorentz factor there stays below $\gamma_{\rm term}$. In our reading of Refs.~\cite{Bodeker:2017cim,Gouttenoire:2021kjv}, the walls of a gauge-singlet field emit no such gauge bosons and run away whenever $\alpha_{\rm r}>\alpha_\infty$. Mixing with the Higgs field would give the gauge bosons a mass change across such walls. A runaway wall in a plasma carries only the fraction $(\alpha_{\rm r}-\alpha_\infty)/\alpha_{\rm r}$ of the released energy~\cite{Espinosa:2010hh}, and the yield, quadratic in the wall energies, falls by the square of this fraction. Equation~\eqref{eq:S-Yuniv} also assumes that the vacuum energy dominates the expansion. The yield of the Letter therefore applies to transitions in vacuum~\cite{An:2026ghw,Freese:2023fcr}, to gauge-singlet wall fields with $\alpha_{\rm r}\gg1$ and $\alpha_{\rm r}\gg\alpha_\infty$, and to plasmas dilute enough that $\gamma_{\rm term}$ exceeds the Lorentz factor at the first collision. Strongly supercooled transitions give exponential nucleation with vacuum domination, and they may realize the benchmark of the Letter if the walls collide before friction stops them. This may happen in parts of the parameter space of a classically scale-invariant $U(1)_{B-L}$ extension of the Standard Model~\cite{Ellis:2019oqb} and in hidden sectors with runaway walls~\cite{Katz:2016adq}. Whether a given model meets these conditions must be checked case by case. 

\section{Realization through Standard-Model vacuum decay}\label{sec:S-anli}

In the scenario of Ref.~\cite{An:2026ghw}, which we call the An--Li scenario, the hot Big Bang begins with the decay of a Standard Model false vacuum of energy density $L$. After inflation, a dark sector dominates and redshifts as $a^{-6}$ (kination), with $a$ the scale factor. Standard Model bubbles nucleate at a constant rate $\Gamma$ per four-volume, and the transition happens when the Hubble rate falls to $H_c=\Gamma^{1/4}$. It lasts about one Hubble time, and the vacuum energy is a fraction $\alpha=L/\rho_{\rm tot}$ of the total. We consider $\alpha=0.05$, $5\times10^{-3}$ and $5\times10^{-4}$, the first at the edge of the bound from primordial black holes in Ref.~\cite{An:2026ghw}.

\emph{Yield.} We compute the yield with a cosmological Monte Carlo on the self-consistent background, in which the Friedmann equation and the conversion of the false vacuum are solved together. The energy of an uncollided wall follows from the vacuum energy released inside its bubble and redshifts as $1/a$~\cite{An:2026ghw}. We write it as $E/A=(L/H_c)\,\hat e_{\rm c}$, with $\hat e_{\rm c}$ a dimensionless integral over the expansion history. Matching it to $E/A=\gamma\sigma_{\rm wall}$ gives $\gamma m/\MP=6\sqrt{3\alpha}\,\hat e_{\rm c}/\nu^2$, independent of $L$. Define an effective inverse duration $\beta_{\rm eff}$ by requiring that the false-vacuum fraction falls from $0.9$ to $0.1$ in the same time as for an exponential rate. On the self-consistent background $\beta_{\rm eff}\approx3.5\,H_c$ for all three values of $\alpha$. With $H$ the Hubble rate of the vacuum energy alone, $\sqrt\alpha\,H_c$, and $\beta=\beta_{\rm eff}$, Eq.~\eqref{eq:S-Yuniv} reproduces the Monte-Carlo normalization to about $10\%$ at fixed latent heat. The DM window uses the Monte-Carlo coefficient itself, known to about $1\%$, and the mass dependence computed on the cosmological first-collision surfaces. We assume that the released energy thermalizes into $g_*=106.75$ degrees of freedom. Thermalization at $T\sim100$~MeV, as estimated in Ref.~\cite{An:2026ghw}, would raise the yield by a factor of about $1.6$, since at fixed latent heat $Y_X\propto g_{*\rho}^{3/4}/g_{*s}$, and $g_{*\rho}\approx17.7$ and $g_{*s}\approx17.3$ at $100$~MeV~\cite{Saikawa:2018rcs}. 

\begin{figure}[t]
\centering
\includegraphics[width=0.9\textwidth]{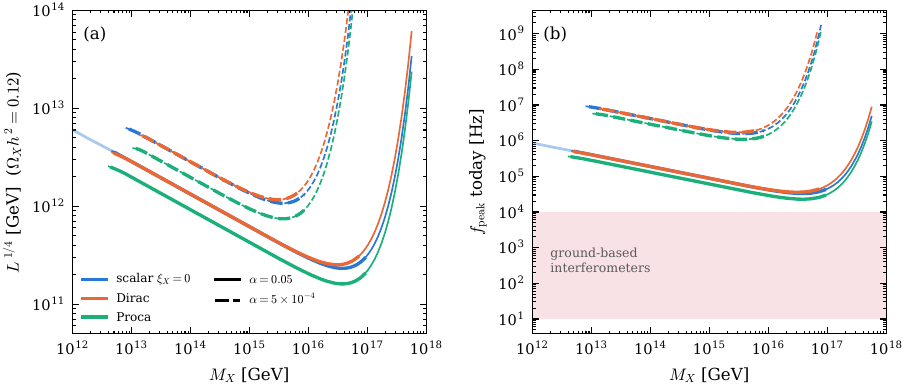}
\caption{(a) Scale of the latent heat $\Lq$ against $M_X$ for $\Omega_Xh^2=0.12$ in the An--Li scenario (kination, $\nu=3$), for minimally coupled scalar, Dirac fermion and Proca vector DM, all with $\xip=0$. Solid curves are for $\alpha=0.05$ and dashed curves for $5\times10^{-4}$. The curves are shown where thermal freeze-in contributes less than $10\%$ of the collision yield. Thin segments have $R<0.1$. The faint low-mass segment of the minimal-scalar curve for $\alpha=0.05$ marks a region where freeze-in is not under control. (b) Present-day GW peak frequency along the same curves, in the bulk-flow model of Ref.~\cite{An:2026ghw}. The shaded band is the frequency range of ground-based interferometers.}
\label{fig:S-anli}
\end{figure}

\emph{Window.} Table~\ref{tab:S-window} lists the window for the Dirac fermion. Over all spins at $\xi_H=0$ and for $\nu\ge2$, the core spans $M_X$ from $4\times10^{12}$ to $2\times10^{17}$~GeV and $\Lq$ from $1\times10^{11}$ to $9\times10^{12}$~GeV (Fig.~\ref{fig:S-anli} shows $\nu=3$). We quote the window for $\nu\ge2$. For $\nu=1$, $28\%$ of the yield at $\alpha=0.05$ comes from collision energies above $\Lambda_{\rm cut}$. In this scenario the wall field is the Higgs boson, so $\xip=\xi_H$. For the Dirac fermion $\xi_H$ matters only near the upper mass edge, which $\xi_H=1$ raises by a factor of about $1.7$.

\begin{table}[t]
\centering
\begin{tabular}{llcccc}
\hline
$\alpha$ & $\nu$ & $M_X$ [GeV] & $\Lq$ [GeV] & $f_{\rm peak}$ [Hz] & $>\Lambda_{\rm cut}$\\
\hline
$0.05$ & $1$ & $6\times10^{12}$--$4\times10^{17}$ & $2\times10^{11}$--$4\times10^{12}$ & $2\times10^4$--$5\times10^5$ & $28\%$\\
$0.05$ & $1.5$ & $5\times10^{12}$--$3\times10^{17}$ & $2\times10^{11}$--$3\times10^{12}$ & $2\times10^4$--$5\times10^5$ & $3\%$\\
$0.05$ & $2$ & $5\times10^{12}$--$2\times10^{17}$ & $2\times10^{11}$--$3\times10^{12}$ & $3\times10^4$--$5\times10^5$ & $0.1\%$\\
$0.05$ & $3$ & $5\times10^{12}$--$7\times10^{16}$ & $3\times10^{11}$--$3\times10^{12}$ & $4\times10^4$--$5\times10^5$ & $0$\\
$5\times10^{-3}$ & $1$ & $8\times10^{12}$--$2\times10^{17}$ & $3\times10^{11}$--$4\times10^{12}$ & $1\times10^5$--$2\times10^6$ & $0.4\%$\\
$5\times10^{-3}$ & $2$ & $8\times10^{12}$--$5\times10^{16}$ & $4\times10^{11}$--$4\times10^{12}$ & $2\times10^5$--$2\times10^6$ & $0$\\
$5\times10^{-3}$ & $3$ & $8\times10^{12}$--$2\times10^{16}$ & $5\times10^{11}$--$4\times10^{12}$ & $2\times10^5$--$2\times10^6$ & $0$\\
$5\times10^{-4}$ & $1$ & $1\times10^{13}$--$7\times10^{16}$ & $6\times10^{11}$--$5\times10^{12}$ & $8\times10^5$--$8\times10^6$ & $0$\\
$5\times10^{-4}$ & $2$ & $1\times10^{13}$--$2\times10^{16}$ & $9\times10^{11}$--$5\times10^{12}$ & $1\times10^6$--$8\times10^6$ & $0$\\
$5\times10^{-4}$ & $3$ & $1\times10^{13}$--$7\times10^{15}$ & $1\times10^{12}$--$5\times10^{12}$ & $2\times10^6$--$8\times10^6$ & $0$\\
\hline
\end{tabular}
\caption{Dark-matter window of the An--Li scenario for a Dirac fermion ($\chat=1$, kination, $\xip=\xi_H=0$), for each transition strength $\alpha$ and field excursion $\nu=\Dphi/\Lq$, with the yield above $\Lambda_{\rm cut}=0.46\,\MP$ discarded. The columns give the range of $M_X$ and of $\Lq$ on the core, the present-day GW peak frequency in the bulk-flow model of Ref.~\cite{An:2026ghw}, and the fraction of the massless yield with collision energy above $\Lambda_{\rm cut}$.}
\label{tab:S-window}
\end{table}

\emph{Conditions on the inflationary era.} The scenario needs a low inflation scale. We computed the gravitational production of $X$ at the end of inflation~\cite{Kolb:2023ydq,Nikandish:2026vpw} with a Bogoliubov calculation on the inflation--kination background~\cite{Chun:2009yu}. It stays below $10\%$ of the DM only if $M_X/H_{\rm inf}\gtrsim70$ to $180$ for an abrupt end of inflation, or $\gtrsim9$ to $21$ for a smooth one. Keeping the Higgs in its false vacuum during $60$ $e$-folds of inflation~\cite{Hawking:1981fz} requires $H_{\rm inf}\lesssim(0.14$--$0.19)\,\Lq$ for $\nu=2$ and $(0.05$--$0.07)\,\Lq$ for $\nu=3$. Friction from radiation produced at the end of inflation~\cite{Bodeker:2017cim} gives the weakest bound. Together, for an abrupt end of inflation, $H_{\rm inf}$ must lie below a bound between $2\times10^{10}$ and $2\times10^{11}$~GeV, depending on $\alpha$ and on the position on the DM curve. The tachyonic mass $-6\xi_HH^2$ during kination must destabilize neither the false vacuum nor the electroweak vacuum afterward~\cite{Opferkuch:2019zbd,Laverda:2024qjt}, which allows $\xi_H\lesssim1$ at the top-quark mass. Black holes formed in the transition~\cite{Fujita:2014hha,Lennon:2017tqq} contribute less than $10^{-10}$ of the DM through Hawking radiation.

\emph{Gravitational waves.} The peak amplitude is $h^2\Omega_{\rm peak}\simeq2.5\times10^{-8}$ for $\alpha=0.05$ and $2.5\times10^{-10}$ for $\alpha=5\times10^{-4}$~\cite{An:2026ghw}. Over the DM window of the Dirac fermion with $\nu\ge2$, the peak frequency lies between $3\times10^4$ and $5\times10^5$~Hz for $\alpha=0.05$ and between $1\times10^6$ and $8\times10^6$~Hz for $\alpha=5\times10^{-4}$, 
 from the bulk-flow model used in Ref.~\cite{An:2026ghw}. The envelope limit would give frequencies about two times higher. Planned detectors reach only the low-frequency tail. For $\alpha=0.05$ the peak lies above the quantum limit of interferometric detection~\cite{Guo:2025cza}, for $f_{\rm peak}\lesssim10^5$~Hz with a photon-counting readout and $f_{\rm peak}\lesssim6\times10^5$~Hz at the channel limit. The background adds $\Delta N_{\rm eff}\approx5\times10^{-3}$ to the effective number of neutrino species for $\alpha=0.05$~\cite{Caprini:2018mtu}, below the current bound~\cite{Planck:2018vyg}.

\section{Beyond the first collision and corrections below the cutoff}\label{sec:S-beyond}

Table~\ref{tab:S-beyond} collects the effects left out of the first-collision yield. All production processes in the first block can only add to the yield. Entropy produced after the transition, for example by long-lived field oscillations, would dilute it. So would a stage in which the inhomogeneous field left by the collisions behaves like matter before it thermalizes~\cite{Mansour:2026sdx}.

After the first collision, the walls continue as energy shells, which cross many times~\cite{An:2026ghw}. Because $\sigma\propto\shat$, any counter-streaming distribution of quanta with energies $\gg M_X$ keeps producing $X$. On the bubble configurations of the cosmological Monte Carlo, freely passing shells raise the yield by $S_{\rm cr}\approx4.3$ for $\alpha=0.05$ and $5.1$ for $\alpha=5\times10^{-4}$ if the hard quanta survive every crossing. The enhancement falls toward one if they degrade within a few crossings, and the degradation rate is not known. Walls in an even potential, as for the Higgs, are trapped and collide again~\cite{Jinno:2019bxw}. We have not computed these re-collisions. During thermalization, graviton exchange between hard quanta and the forming bath adds up to about $20\%$ ($\alpha=0.05$) and $80\%$ ($\alpha=5\times10^{-4}$) for $M_X\lesssim10^{15}$~GeV, an estimate uncertain by a factor of a few.

\begin{table}[t]
\centering
\begin{tabular}{lll}
\hline
effect & size & sign\\
\hline
later shell crossings, free passage & $S_{\rm cr}\approx4.3$ ($\alpha=0.05$), $5.1$ ($\alpha=5\times10^{-4}$) & adds\\
re-collisions of trapped walls & not computed & adds\\
pre-thermal hard--soft scatterings & up to $20\%$ to $80\%$ ($M_X\lesssim10^{15}$~GeV) & adds\\
late thermalization at $T\sim100$~MeV (An--Li) & factor $\approx1.6$ & adds\\
\hline
expansion during the transition & factor $\approx e^{-4H/\beta}$ ($0.67$ at $H/\beta=0.1$) & lowers\\
loop corrections below $\Lambda_{\rm cut}$ & at most $16\%$ times a logarithm, near the upper mass edges & either\\
second-order contact terms & $\lesssim10^{-2}$ of the yield & either\\
strong-gravity effects & suppressed by $(H/\beta)^2$ or $\alpha$ & either\\
black-hole formation ($0.01\le H/\beta\le0.1$) & collapse probability $<10^{-40}$ & none\\
\hline
\end{tabular}
\caption{Effects not included in the first-collision yield. The first block lists production processes, the second corrections to the first-collision yield itself.}
\label{tab:S-beyond}
\end{table}

Loop corrections are estimated conservatively as $N\shat/(16\pi^2\MP^2)$ times a logarithm. Below the cut, this is at most $N\Lambda_{\rm cut}^2/(16\pi^2\MP^2)\approx16\%$, reached only near the upper mass edges. Second-order contact terms, the $XXhh$ seagull and three-graviton diagrams in which the field of each wall enters once, are suppressed relative to the leading amplitude by $[\gamma^2m^2\Dphi^2/(\MP^2p_1p_2)]^2$. Integrated, they give $\sim(\Dphi/\MP)^4(M_{\rm coll}/M_X)^2\ln(M_{\rm coll}/M_X)$, which is $\lesssim10^{-10}$ of the yield in the realization of Sec.~\ref{sec:S-anli} and $\lesssim10^{-2}$ at the light end of the benchmark cores. Near threshold at the lightest masses the amplitude ratio exceeds one, so this power counting fails there. The large ratio comes from the coherent soft partons of the other wall. Summed, they form the classical metric of that wall, and its strength over a parton wavelength, $(E/A)/(\MP^2p)\lesssim10^{-2}$, then measures the correction. This suggests that the correction stays small, but an explicit calculation remains to be done. 
 The collisions reach neither black-hole formation, which needs $\sqrt{\shat}\gg M_{\rm Pl}$, nor the eikonal regime $G\shat\gg1$. The two sheets focus each other only at a distance $\simeq2\MP^2/(E/A)$, which exceeds $\beta^{-1}$ by $\simeq2\beta^2/(3H^2)\gg1$, and shock-wave deflection is suppressed by $(H/\beta)^2$ in the benchmark and by $\alpha$ in Sec.~\ref{sec:S-anli}. An $\mathcal R^2$ term in the gravitational action affects only the trace channel. A Weyl-squared term modifies spin-2 exchange at momentum transfers $q^2\sim\MP^2/\kappa_W$, with $\kappa_W$ its dimensionless coefficient. The only contact-type ambiguity that is not suppressed is the curvature coupling $\xip$ of the wall field (Sec.~\ref{sec:S-xsec}). 
 Three tests remain for nonlinear simulations. The first is the oblique factor $\eperp^3$, of which only the flux factor was checked numerically. The second is the effect of the gauge and Yukawa interactions of a Higgs wall during the overlap. The third is free passage for walls accelerated from rest.

\end{document}